# On the Relationship Between Coupling and Refactoring: An Empirical Viewpoint

Steve Counsell
Dept. of Computer Science
Brunel University
London, UK
steve.counsell@brunel.ac.uk

Mahir Arzoky
Dept. of Computer Science
Brunel University
London, UK
mahir.arzoky@brunel.ac.uk

Giuseppe Destefanis
Dept. of Computer Science
Brunel University
London, UK
giuseppe.destefanis@brunel.ac.uk

Davide Taibi
Tampere University
Tampere, Finland
davide.taibi@tuni.fi

*Abstract*—[Background] Refactoring has matured over the past twenty years to become part of a developer's toolkit. However, many fundamental research questions still remain largely unexplored. [Aim] The goal of this paper is to investigate the highest and lowest quartile of refactoring-based data using two coupling metrics – the Coupling between Objects metric and the more recent Conceptual Coupling between Classes metric to answer this question. Can refactoring trends and patterns be identified based on the level of class coupling? [Method] In this paper, we analyze over six thousand refactoring operations drawn from releases of three open-source systems to address one such question. [Results] Results showed no meaningful difference in the types of refactoring applied across either lower or upper quartile of coupling for both metrics; refactorings usually associated with coupling removal were actually more numerous in the lower quartile in some cases. A lack of inheritance-related refactorings across all systems was also noted. [Conclusions] The emerging message (and a perplexing one) is that developers seem to be largely indifferent to classes with high coupling when it comes to refactoring types – they treat classes with relatively low coupling in almost the same way.

## I. INTRODUCTION

Over the past twenty or so years, refactoring has become a mainstream software engineering discipline; hundreds of empirical studies have been undertaken since Fowler's seminal text and Opdyke's preliminary study were first published [7, 10]. In short, refactoring is: "The process of changing a software system in such a way that it does not alter the external behavior of the code yet improves its internal structure." While previous studies have given us a broad understanding of the area, there are still many aspects to refactoring that we still do not appreciate. For example, it is intuitive to suggest that the extract method refactoring (which splits one method into two or more methods) might be applied to methods exhibiting the long method code smell [7]. Equally, the extract class refactoring which splits one class into two or more separate classes might reasonably be applied to large classes [7]. What is not so clear, however, is when the motivation for specific refactorings is less intuitive.

In this paper, we explore a large dataset of refactorings applied to multiple releases of three open-source systems. We used data produced by Bavota et al. [3], as a basis of our analysis – the same dataset is freely downloadable for replication and further study from their original paper. We explore those refactorings with specific reference to coupling. Excessive coupling is generally considered harmful and it is widely accepted in the software engineering community that developers should strive to keep coupling to a minimum [12]. We focus on two specific metrics: firstly, the well-used Coupling between Objects metric (CBO) of Chidamber and Kemerer [6] which counts the number of classes to which any single class is coupled. The second is the Conceptual Coupling between Classes metric (CCBC) developed by Poshyvanyk et al. [11], and is a measure of the textual similarity between classes. The higher that textual similarity, the higher the coupling value of the metric. To further facilitate our analysis, we decomposed the same dataset taken from [3] into inter-quartile ranges. This gave us three parts to our analysis: the lower 25% of coupling values, the 50% in the mid-range of coupling values (which, for the purposes of this paper where we look at the upper and lower quartiles, we henceforth ignore) and the upper 25% of coupling values. The underlying premise of the research question is that there will generally be fewer refactorings in the lower quartile of classes compared to the upper quartile when ranked on coupling and, also, we will find disjoint sets of refactoring types applied in classes with low coupling, vis a vis classes with high coupling.

Results showed no very little difference in the types of refactoring applied across either quartile (i.e., we found a very high overlap of refactoring types) – this applied to both metrics studied; secondly, refactorings usually associated with coupling removal were found to be more numerous in the lower quartile, in some cases. Finally, very few inheritance-related refactorings were found across all systems. These results present a situation which is difficult to explain and calls into question our belief about high coupling and its corrosive influence; certainly, if developer refactoring is anything to go by. Also, we need to fundamentally question the value of using Fowler's complete set of seventy-two refactorings [7] since only a handful of that set seem to ever be applied (and certainly not inheritance-based ones). The remainder of the paper is organized as follows. In the next section, we describe preliminary information on the systems studied, the data collected and summary data. We then present results through an analysis of each the three open-source systems (Section 3 using CBO and CCBC. Section 4 discusses related work; threats to study validity are then reported in Section 5. Finally, we conclude and point to further work (Section 6).

## II. PRELIMINARIES

The systems studied consisted of three Java open source projects: Xerces [17], Apache Ant [15] and ArgoUML [16]. Xerces-J is a Java XML parser, Apache Ant a build tool and library primarily designed for Java applications and ArgoUML a UML modeling tool and Table 1, taken verbatim from [3] shows the salient characteristics of the three systems. Here, 'Rel.' is the number of releases and the final column (#Ref.) represents the total number of refactorings for that system before we decomposed it into its inter-quartile ranges.

Table 1. Features of the three systems analyzed from [3]

| System | Period | Analyzed | Rel. | # Classes | # Ref. |
|---|---|---|---|---|---|
| Xerces | Nov '99-Nov '10 | 1.0.4-2.9.1 | 33 | 181-776 | 7502 |
| Apache Ant | Jan '00-Dec '10 | 1.2-1.8.2 | 17 | 87-1191 | 1289 |
| ArgoUML | Oct '02-Dec '11 | 0.12-0.34 | 13 | 777-1519 | 3255 |

Table 2 shows data the number of refactorings in each of the ranges for the three systems after it had been decomposed. It shows the Upper Quartile (UQ) and Lower Quartile (LQ) median values for the CBO and CCBC metrics and the Inter-Quartile Range (IQR) in each case. IQR is calculated as the UQ value minus the LQ value. It also shows the number of refactorings in each of the lower and upper ranges in parentheses after each value. For example, in the UQ of the Xerces systems, the median was 26 and the number of refactorings 1818. The IQR for the CBO was 21 and 8.1 for the CCBC metric.

Table 2. Dataset decomposition into quartiles

| System | CBO (UQ) | CBO (LQ) | IQR | CCBC (UQ) | CCBC (LQ) | IQR |
|---|---|---|---|---|---|---|
| Xerces | 26 (1818) | 5 (1745) | 21 | 3.15 (1875) | 11.25 (1667) | 8.1 |
| Apache | 9 (288) | 2 (121) | 7 | 9.65 (316) | 0.86 (289) | 8.79 |
| ArgoUML | 17 (627) | 4 (601) | 13 | 19.44 (813) | 3.90 (827) | 15.54 |

A total of 5200 refactorings were therefore used as a basis for our analysis in the LQ and UQ data (the sum of the values in parentheses in Table 2). Finally, the Ref-Finder tool [8] was used to extract the set of refactorings on which our analysis was based; the refactorings were extracted and validated as part of the earlier study and are used in our study [3]. The tool collects up to sixty-three of Fowler's original set of 72 refactorings. Ref-Finder has a recall of 95% and a precision of 79% [8].

## III. RESULTS

We begin by looking at the CBO metric and the refactorings in those two quartiles, before moving on to the CCBC.

### A. CBO refactorings applied

Figs. 1a-1f show the distribution of the top ten most popular refactorings for the CBO metric in the Xerces, Apache Ant and ArgoUML systems for the lower quartile ranked in ascending order (Fig. 1a, 1c and 1e) and the UQ ranked similarly (Fig. 1b, 1d and 1f). We chose the top ten refactorings simply for ease of comparison between figures and we report the totals in that analysis where appropriate. The refactoring acronyms in the figures are as follows: RM: Rename method; MF: Move Field; MM: Move Method; RP: Remove Parameter; CCE: Consolidate Conditional Expression; AP: Add Parameter; RMNwSC: Replace Magic Number with Symbolic Constant; CDCF: Consolidate Duplicate Conditional Fragments; IEV: Introduce Explaining Variable; RMwMO: Replace Method with Method Object; EM: Extract Method; RCF: Remove Control Flag. The frequency of each refactoring is on the x-axis.

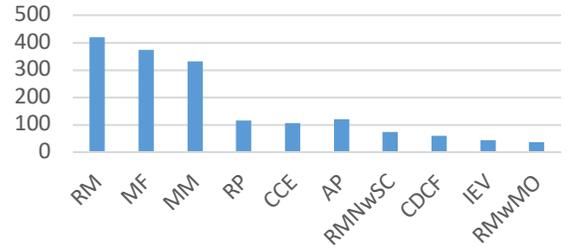

Fig. 1a: Xerces LQ

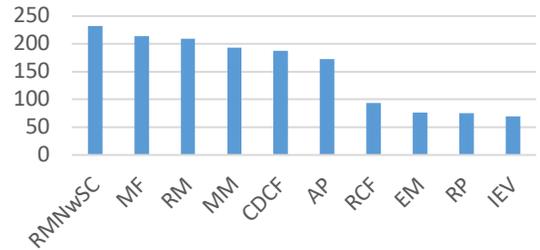

Fig. 1b: Xerces UQ

One noticeable feature from Figs. 1a and 1b is the overlap in the types of refactoring applied in each quartile. The MF, RM and MM refactorings all feature in the top four in each case. In fact, of the ten refactorings in each figure, seven are common to both. It is also interesting that the LQ and the UQ both had large numbers of MM and MF refactorings. These are refactorings which are strongly associated with coupling reduction, since they move features between classes to where those features are most needed. The motivation for MM is when [14]: "*A method is used more in another class than in its own class*". The solution is to: "*Create a new method in the class that uses the method the most, then move code from the old method to there.*" A similar motivation and solution applies to

MF. It was therefore surprising to see so many of each refactoring type in both quartiles. In the LQ, approximately 25.67% of all refactorings were attributed to just MM and MF. This contrasts with only 37.54% for the UQ; in other words, a comparable percentage of these coupling-related refactorings were found to have been applied in lowly-coupled classes. Even more notable is that only fifteen and sixteen of Fowler's original refactorings can be found across the UQ and LQ, respectively from the sixty-three refactorings that Ref-Finder extracts. In other words, the appropriateness, relevance of many of Fowler's original and complete set of refactorings could be questioned in this case. Figs. 1c and 1d for Apache show a similar pattern to the Xerces system. Six of the refactorings are common across the set of ten refactorings in the figures.

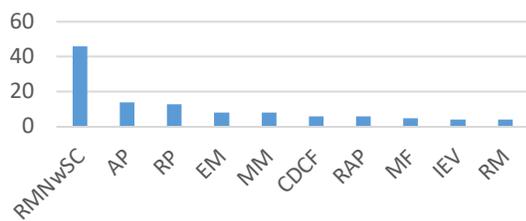

Fig. 1c: Apache LQ

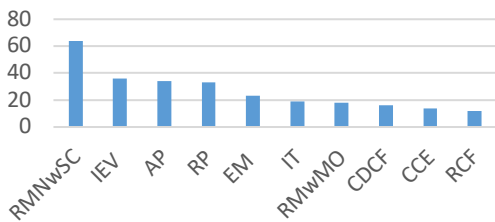

Fig. 1d: Apache UQ

One standout feature from both figures is the proportion of RMNwSC refactorings in both the LQ and UQ. This refactoring accounted for over 13.15% and 20.25% of the total number of refactorings in the LQ and UP, respectively. The RMNwSC is not, strictly speaking, a refactoring which is directly associated with coupling. It is one that simply replaces a literal with a constant. For example, it replaces all hard-coded values of 3.142 in the code with a constant let us say 'Pi'. Again, as for the Xerces system, the percentage of MM and MF refactorings was comparable in the LQ and UQ. In fact MM accounted for 12.11% of refactorings in the LQ compared with just 3.16% in the UQ. What is also noticeable is the relatively large number of AP and RP refactorings in each quartile. The motivation for using AP is when [14]: "*A method doesn't have enough data to perform certain actions*". The solution is to: "*Create a new parameter to pass the necessary data*". In the LQ, these two refactorings accounted for 21.19% of the total and in the UQ it was 17.64%. One hypothesis as to why this might have been the case is that sharing of data and methods, usually managed by moving class features around might have instead been accomplished through addition and removal of parameters to the method signatures and accessing functionality that way - obviating the need to move features around. Only sixteen of Fowler's seventy-two refactorings were identified across both quartiles. Finally, Figs. 1e and 1f show the data for the ArgoUML system and these share eight refactorings from the ten shown in each figure. The number of MM and MF in the LQ accounted for approximately 35.17% of the total refactorings. Thus contrasts with just 10.39% for the UQ. Again, the AP and RP refactorings figure quite strongly. The RMwMO is the most common refactoring in the LQ (applied 99 times).

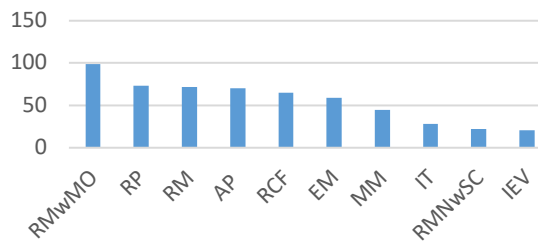

Fig. 1e: ArgoUML LQ

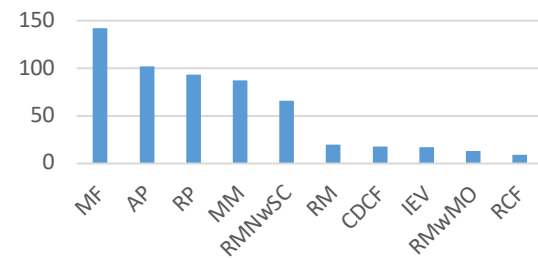

Fig. 1f: ArgoUML UQ

The motivation for using the RMwMO refactoring is when [14]: "*You have a long method in which the local variables are so intertwined that you can't apply Extract Method*". The mechanics of this refactoring are: "*Transform the method into a separate class so that the local variables become fields of the class. Then you can split the method into several methods within the same class*". Here is a refactoring which, far from reducing coupling, actually creates it, since a new class is extracted as part of the process. Only twenty of Fowler's refactorings were found to have been applied across the two quartiles. It was also remarkable from the data how few inheritance-related refactorings had been applied across *all* three systems. In the six quartiles (LQ and UQ for each of Xerces, Apache and ArgoUML, only 112 out of 5200 (2.15%) were related to inheritance; 106 of those were in the Xerces system, primarily consisting of the Push down field (41) and Push down method (43) refactorings. These two refactorings move fields and methods from super classes to their subclasses.

> Considering the data from the CBO, there is a clear correspondence and overlap between refactorings applied in each quartile. Refactorings that we normally associate with removal of high coupling (Move method and Move field) sometimes appeared more frequently in the lower quartile of the data.

## B. CCBC refactorings applied

Just as we analyzed the CBO metric in the previous section, we next compare the results we obtained from that metric and the CCBC, since they compute coupling in contrasting ways. The CBO metric is simply a count of the number of other classes to which a specific class is coupled; the CCBC measures the textual similarity of tokens between classes as a measure. Figs. 2a-2f shows the same data and in the same format shown for the CBO metric, but this time for CCBC. Figs 2a, 2c and 2e represent the UQ for each of the three systems and Figs 2b, 2d, and 2f the LQ. Figs 2a and 2b show a similar pattern to that for the CBO. In fact, nine out of the ten refactorings in Fig 2a are the same as in Fig 1a, although the ordering is slightly different. Similarly, Fig 2b has all ten corresponding refactorings to that found for CBO. The same types of refactoring dominate once more as they did for CBO. The MF, RM and MM feature strongly in the UQ, but they feature heavily in the LQ also. The prominence of these three refactorings is such that 35.38% of the total number of refactorings in the LQ can be attributed to just these three and account for 59.88% in the UQ.

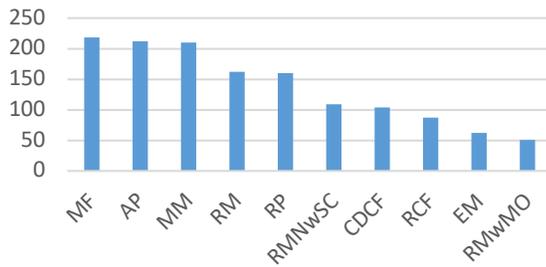

Fig. 2a: Xerces LQ

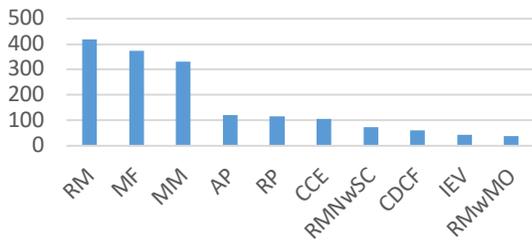

Fig. 2b: Xerces UQ

Figs 2c and 2d show the same data, but for the Apache system. Here, the most prominent refactoring in the UQ was the RM refactoring. For the UQ, RMNwSC was the most popular. This refactoring has very little relationship to coupling (as previously described). The same could be said of many of the refactorings in Figs 2c and 2d and for the other two systems also. For example, Introduce Explaining Variable (IEV) simply replaces an expression with sub-parts assigned to more meaningful variable names. The Consolidate Duplicate Conditional Fragments (CDCF) refactoring is, again, not strictly linked to coupling reduction, nor too Consolidate Conditional Expression (CCE) or Introduce Explaining Variable (IEV). Put another way, many of the refactorings from Fig 2b are not directed at coupling *per se*.

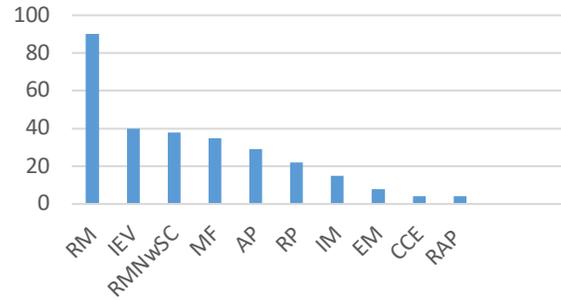

Fig. 2c: Apache LQ

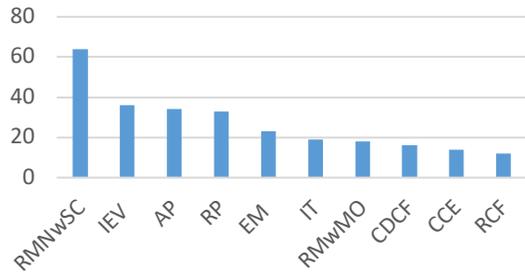

Fig. 2d: Apache UQ

Finally, Figs 2e and 2f show the data for the ArgoUML system. The same types of refactoring as found for the previous two systems recur. The most popular refactoring in the LQ is AP (130), closely followed by RP (128). For the UQ, MF is the most popular (198) followed by RMwMO (134). Seven of the ten refactorings are common to both figures, emphasizing again the overlap between the two quartiles.

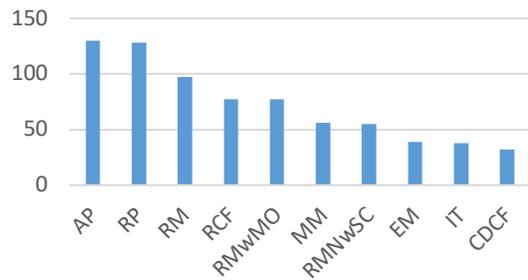

Fig. 2e: ArgoUML LQ

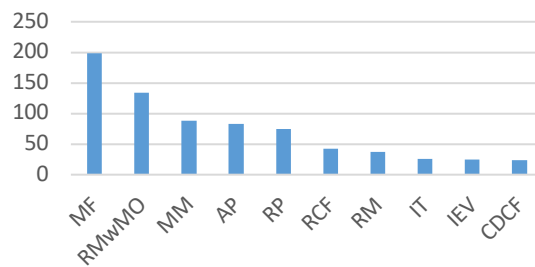

Fig. 2f: ArgoUML (UQ)

Once again, refactorings such as RCF, RMNwSC and CDCF feature heavily. For this system again only twenty of Fowler's refactorings were identified and many of those were just in single figures.

*C. The influence of size*

The results so far for both CBO and CCBC metrics have shown that the upper and lower quartiles comprise large numbers of MM, MF, AP and RP to name several. However, the implicit assumption we have made is that the LQ of the CBO and CCBC will tend to be for smaller classes than those in the upper quartile. Table 3 shows the minimum, maximum, mean and median number of lines of code (LOC) and methods per class (WMC) for just the CBO metric in the LQ and UQ, for each of the three systems. So, for example for Xerces, the mean number of LOC in the LQ was 249.46. The mean LOC in the UQ was 1781.36. We can see that for all systems, the discrepancy between the UQ and LQ is large. For every measure in Table 3, the UQ value exceeds LQ. Consider ArgoUML. The mean WMC in the LQ was 15.00; in the UQ it was 57.72. The data in Table 3 thus implies that even though classes in the UQ were much larger than classes in the LQ, comparable numbers and types of refactorings were applied in each.

Table 3. Summary statistics for size (all systems)

| Metric/Stat. | Min. | Max. | Mean | Med. |
|---|---|---|---|---|
| **Xerces** | | | | |
| LOC (LQ) | 12 | 2503 | 249.46 | 158 |
| LOC (UQ) | 160 | 5827 | 1781.36 | 1832 |
| WMC (LQ) | 0 | 243 | 45.81 | 38 |
| WMC (UQ) | 5 | 687 | 221.33 | 224 |
| **Apache Ant** | | | | |
| LOC (LQ) | 9 | 320 | 120.03 | 102 |
| LOC (UQ) | 79 | 1973 | 970.83 | 1073 |
| WMC (LQ) | 2 | 58 | 23.34 | 18 |
| WMC (UQ) | 23 | 371 | 166.82 | 107 |
| **ArgoUML** | | | | |
| LOC (LQ) | 8 | 357 | 62.40 | 40 |
| LOC (UQ) | 35 | 1424 | 433.78 | 317 |
| WMC (LQ) | 0 | 93 | 15.00 | 7 |
| WMC (UQ) | 2 | 252 | 57.72 | 42 |

## IV. RELEVANT RELATED WORK

Refactoring has been the subject of multiple empirical studies in the past. In this paper, we have focused on two coupling metrics both of which have both been studied before [4, 5, 11, 13]; work by Bavota et al., has shown the CCBC to be the coupling metric which captures a developer's perception of coupling between code components best [2]. However, the CBO still effectively remains the 'gold' standard for measuring coupling and used in hundreds of studies of code in the past [1, 4]. The work in this paper suggests that from a refactoring perspective, the two metrics are very similar in terms of how they relate to high and low coupling. The data for our analysis was used first by Bavota et al., [3] in a study of the relationship between refactoring and quality through a set of code quality metrics. The paper mined the evolution history of the same three Java open source projects and investigated whether refactoring activities applied to code suggested refactoring might be necessary. Results indicated the metrics did not show a clear relationship with refactoring. In their own words: "*….refactoring operations are generally focused on code components for which quality metrics do not suggest there might be need for refactoring operations.*" In fact, in [3], the CBO was found to be related most strongly to the Introduce null object, Pull up field, Push down method and Replace data with object refactorings and not the refactorings we'd expect (i.e., Move method, Move field, etc). Equally, the CCBC was found to be related to the Separate query from modifier refactoring and not the similar expected refactorings. Our results suggest that perhaps in highly- and lowly-coupled classes these relationships might not hold. In terms of inheritance and why we found so low numbers of refactorings, it has long been acknowledged that developers do not use it to the depth we might expect [1]. Perhaps the lack of inheritance related refactorings is simply due to the complexity involved in manipulating inheritance structures. Finally, a review of refactoring was undertaken in the early 00's [9].

## V. THREATS TO VALIDITY

The preceding analysis raises a number of questions about how we view refactoring. In terms of whether these results are 'emerging', we claim that the themes of the paper (i.e., lack of use of most of Fowler's refactorings, overlap in refactoring usage and our understanding of contrasting coupling metrics) come together to form emerging and pressing research issues. In terms of threats to validity of the study, firstly, we have only used three systems as the basis of our analysis. However, we see the work as largely exploratory and represent the results from a sample set of open-source systems. Secondly, we have looked at three different open-source domains and so generalizing the results to other domains may prove difficult as a result. Thirdly, perhaps we are wrong to think about coupling in isolation as a class factor to consider. Many of the refactorings such as CCDF and RMNwSC which we found to be applied frequently address *complexity* in code other than coupling. Fourthly, we have excluded the mid-range refactoring data from our analysis (we considered upper and lower quartiles only). Analysing this data may shed light on our results.

Another threat is the selection of CBO and CCBC as coupling metrics. Other coupling metrics, such as the ones defined in [2],[4],[18], could have led to different results.
Finally we have not considered the role of defects in our analysis. It may simply be that high coupling in a class is necessary and actually leads to stable classes, relatively free of defects.

## VI. CONCLUSIONS AND FURTHER WORK

Until now, the refactoring research community has predominantly focused its efforts in refactoring studies on Fowler's set of seventy-two refactorings [7], but the relevance of the *complete* set has rarely been questioned. In this paper, the notion that highly coupled classes will be refactored in a different way to lowly-coupled classes is explored through the prism of refactoring. A set of refactoring data from three open-source systems sought to answer one question: can refactoring trends and patterns be identified based on the level of class coupling? We assumed, rightly or wrongly, that developers would target classes with high coupling, and would use the relevant set of refactorings to do so. Results showed no obvious difference in the types of refactoring applied across either lowly- or highly-coupled classes, suggesting that developers do not tend to adhere to this principle. In fact, refactorings usually associated with coupling removal were actually found in some cases to be more numerous in low-coupled classes. A dearth of inheritance-related refactorings across all systems was also noted and although the lack of such refactorings has been noted in other studies [1], no concrete evidence for why this is the case has been put forward. Finally, the CBO and CCBC metrics showed a very strong relationship, suggesting that they may be surrogates (further work could examine their inter-relationship more closely).

There are many other avenues for further work. Firstly, replication of this study on a large-scale would be a valuable way of confirming or refuting the reported results. We are planning to replicate this work starting from the Technical Debt Dataset, that include refactoring information on 33 projects. Only a small number of the seventy-two refactorings of Fowler seem to be used "in anger" and this begs the question whether other more relevant and appropriate refactorings are hidden in the code that we are ignoring? Secondly, the work has looked at the two coupling metrics, without considering class cohesion [12]. It may be useful and insightful to consider this aspect of code in conjunction with coupling. Finally, the same empirical study could be carried out using proprietary code, complementing the open-source work in this paper. The summary refactoring data used in this paper can be made available from the lead author.